\documentclass{article}
\usepackage{amsmath, amssymb, amsfonts}
\title{On a regular modified Schwarzschild spacetime}  
\author{Hristu Culetu, \\Ovidius University, Dept.of Physics, \\ Mamaia Avenue 124, 900527 Constanta, Romania, \\e-mail : hculetu@yahoo.com}

\begin{document}
\numberwithin{equation}{section}
\pagenumbering{arabic}
\maketitle
\newcommand{\fv}{\boldsymbol{f}}
\newcommand{\tv}{\boldsymbol{t}}
\newcommand{\gv}{\boldsymbol{g}}
\newcommand{\OV}{\boldsymbol{O}}
\newcommand{\wv}{\boldsymbol{w}}
\newcommand{\WV}{\boldsymbol{W}}
\newcommand{\NV}{\boldsymbol{N}}
\newcommand{\hv}{\boldsymbol{h}}
\newcommand{\yv}{\boldsymbol{y}}
\newcommand{\RE}{\textrm{Re}}
\newcommand{\IM}{\textrm{Im}}
\newcommand{\rot}{\textrm{rot}}
\newcommand{\dv}{\boldsymbol{d}}
\newcommand{\grad}{\textrm{grad}}
\newcommand{\Tr}{\textrm{Tr}}
\newcommand{\ua}{\uparrow}
\newcommand{\da}{\downarrow}
\newcommand{\ct}{\textrm{const}}
\newcommand{\xv}{\boldsymbol{x}}
\newcommand{\mv}{\boldsymbol{m}}
\newcommand{\rv}{\boldsymbol{r}}
\newcommand{\kv}{\boldsymbol{k}}
\newcommand{\VE}{\boldsymbol{V}}
\newcommand{\sv}{\boldsymbol{s}}
\newcommand{\RV}{\boldsymbol{R}}
\newcommand{\pv}{\boldsymbol{p}}
\newcommand{\PV}{\boldsymbol{P}}
\newcommand{\EV}{\boldsymbol{E}}
\newcommand{\DV}{\boldsymbol{D}}
\newcommand{\BV}{\boldsymbol{B}}
\newcommand{\HV}{\boldsymbol{H}}
\newcommand{\MV}{\boldsymbol{M}}
\newcommand{\be}{\begin{equation}}
\newcommand{\ee}{\end{equation}}
\newcommand{\ba}{\begin{eqnarray}}
\newcommand{\ea}{\end{eqnarray}}
\newcommand{\bq}{\begin{eqnarray*}}
\newcommand{\eq}{\end{eqnarray*}}
\newcommand{\pa}{\partial}
\newcommand{\f}{\frac}
\newcommand{\FV}{\boldsymbol{F}}
\newcommand{\ve}{\boldsymbol{v}}
\newcommand{\AV}{\boldsymbol{A}}
\newcommand{\jv}{\boldsymbol{j}}
\newcommand{\LV}{\boldsymbol{L}}
\newcommand{\SV}{\boldsymbol{S}}
\newcommand{\av}{\boldsymbol{a}}
\newcommand{\qv}{\boldsymbol{q}}
\newcommand{\QV}{\boldsymbol{Q}}
\newcommand{\ev}{\boldsymbol{e}}
\newcommand{\uv}{\boldsymbol{u}}
\newcommand{\KV}{\boldsymbol{K}}
\newcommand{\ro}{\boldsymbol{\rho}}
\newcommand{\si}{\boldsymbol{\sigma}}
\newcommand{\thv}{\boldsymbol{\theta}}
\newcommand{\bv}{\boldsymbol{b}}
\newcommand{\JV}{\boldsymbol{J}}
\newcommand{\nv}{\boldsymbol{n}}
\newcommand{\lv}{\boldsymbol{l}}
\newcommand{\om}{\boldsymbol{\omega}}
\newcommand{\Om}{\boldsymbol{\Omega}}
\newcommand{\Piv}{\boldsymbol{\Pi}}
\newcommand{\UV}{\boldsymbol{U}}
\newcommand{\iv}{\boldsymbol{i}}
\newcommand{\nuv}{\boldsymbol{\nu}}
\newcommand{\muv}{\boldsymbol{\mu}}
\newcommand{\lm}{\boldsymbol{\lambda}}
\newcommand{\Lm}{\boldsymbol{\Lambda}}
\newcommand{\opsi}{\overline{\psi}}
\renewcommand{\tan}{\textrm{tg}}
\renewcommand{\cot}{\textrm{ctg}}
\renewcommand{\sinh}{\textrm{sh}}
\renewcommand{\cosh}{\textrm{ch}}
\renewcommand{\tanh}{\textrm{th}}
\renewcommand{\coth}{\textrm{cth}}

\begin{abstract}
A modified version of the Schwarzschild geometry is proposed. The source of curvature comes from an anisotropic fluid with $p_{r} = -\rho$ and fluctuating tangential pressures. The event horizon has zero surface gravity but the invariant acceleration of a static observer is finite there. The Tolman - Komar energy of the gravitational fluid changes sign on the horizon, equals the black hole mass at infinity and is vanishing at $r = 0$. There is a nonzero surface stress tensor on the horizon due to a jump of the extrinsic curvature. Near-horizon geometry resembles the Robinson - Bertotti metric and not the Rindler metric which corresponds to the standard Schwarzschild spacetime. The modified metric has no signature flip when the horizon is crossed and all the physical parameters are finite throughout the spacetime. 
 \end{abstract}
 
 \section{Introduction}
  The Schwarzschild (KS) spacetime is the exact spherically symmetric vacuum solution of Einstein's equations. It is expectable that quantum effects will play a main role in the late stages of the gravitational collapse and the subsequently Hawking evaporation process. 
	
	At the classical level, Dymnikova \cite{ID} (see also \cite{DK}) proposed a nonsingular black hole (BH) solution of Einstein's equations which for large $r$ coincides with the KS spacetime but for small $r$ resembles the de Sitter (deS) spacetime, sourced by a stress tensor with $T^{0}_{~0} = T^{1}_{~1}$ and $T^{2}_{~2} = T^{3}_{~3}$, so that it has an infinite set of comoving reference frames and may therefore describe the spherically symmetric vacuum.
  
 Bonanno and Reuter \cite{BR} altered the KS metric using a running Newton constant and studied the quantum gravitational effects on the dynamics of geometry. In their view, the black hole evaporation stops when its mass approaches some extremal (critical) value $M_{cr} \approx m_{P}$, where $m_{P}$ is the Planck mass. Consequently, a ''cold'' soliton-like remnant is formed so that the classical singularity at $r = 0$ is removed. According to the authors of \cite{BR}, that quantum BH with $M = M_{cr}$ (corresponding to the extremal charged Reissner - Nordstrom BH) may be considered the final state of the KS black hole evaporation process. 
 
 The non-commutativity of spacetime related to the behavior of radiating KS black holes has been put forward by Nicolini et al. \cite{NSS}. Their conclusion is that the evaporation end-point is a zero-temperature extremal BH with no singularity at the origin and with a regular de Sitter (deS) core close to $r = 0$. Their model includes an anisotropic fluid with radial pressure $p_{r} = - \rho$ which is not the common inward pressure of outer layers of matter on the core of a star but a different kind of ''quantum'' outward push induced by non-commuting coordinate quantum fluctuations. In addition, Nicolini \cite{PN} showed that the Gaussian de-localization of a gravitational source in the noncommutative spacetime eliminates the divergent behavior of Hawking temperature. There is a maximal temperature that the BH can reach before cooling down to absolute zero. In his view, noncommutativity regularizes divergent quantities in the final stage of BH evaporation.
 
 Myung et al. \cite{MKP} and Hayward \cite{SAH} introduced the so-called ''regular'' black holes (RBHs) that avoids the curvature singularity beyond the event horizon. The singularity is replaced by a deS spacetime. While Myung et al. detailed their study to universal thermodynamic properties of RBHs, Hayward has discussed the formation and evaporation process of a RBH with a minimal size which can be related with the minimal length induced by string theory.
 
 Recently Xiang et al. \cite{XLS} showed that a BH would not evaporate entirely but will approach a Planck order remnant. They found a finite Kretschmann scalar for the modified KS geometry and a bounded BH mass due to the existence of the horizon. If the BH curvature singularity is removed by quantum gravitational effects, it would cease evaporation at a specific mass scale.
 
 We suggest in this letter a slight modification of the Xiang et al. gravitational potential, coming out in this way physical parameters in closed form. Our model is purely classical and the effective anisotropic gravitational fluid seems not to simulate any quantum effects. The modified KS spacetime proposed has only one horizon and is free of any singularity at $r = 0$. We compute the components of the stress tensor which acts as the source of curvature and depends on the BH mass. We then give the expressions of the ADM mass and of the Tolman - Komar gravitational energy, which vanishes at the horizon and equals BH mass at infinity. All dynamical parameters are finite everywhere. \\
 Throughout the paper the geometrical units $G = c = 1$ are used, excepting numerical calculations where $G$ appears explicitly.
 
 \section{Modified Schwarzschild metric}
Let us consider firstly the Xiang et al. \cite{XLS} modified form of the KS classical metric
   \begin{equation}
  ds^{2} = -(1 + 2\Phi) dt^{2} + (1 + 2\Phi)^{-1} dr^{2} + r^{2} d \Omega^{2}, 
 \label{2.1}
 \end{equation}
 where $\Phi(r) = -(m/r) exp(-\epsilon(r))$ (with $\epsilon(r) > 0)$ and $d \Omega^{2}$ stands for the metric on the unit 2-sphere. The unknown function $\epsilon(r)$ is a damping factor needed to remove the singularity when $r \rightarrow 0$ \cite{XLS}. 
 
 Our main assumption is to consider $\Phi(r)$ to be of the form
    \begin{equation}
    \Phi(r) = - \frac{m}{r} e^{-\frac{k}{r}}
 \label{2.2}
 \end{equation}
  where $k$ is a positive constant that will be chosen such that the equation $g_{tt} = 0$ to have one zero only. Plugging $x = 1/r$ in $1 + 2\Phi = 0$, one obtains
     \begin{equation}
     e^{kx} = 2mx.
 \label{2.3}
 \end{equation}  
 It is clear that (2.3) has one solution when the straightline $2mx$ is tangent to the exponential curve $e^{kx}$ at, say, $x_{H}$. In other words, the two functions and their first derivative are equal at $x_{H}$. One obtains $x_{H} = 1/k = e/2m$, with $lne = 1$. Therefore, the metric (2.1) becomes in our conditions
     \begin{equation}
   ds^{2} = -\left(1 - \frac{2m}{r} e^{-\frac{2m}{er}}\right) dt^{2} + \frac{1}{1 - \frac{2m}{r} e^{-\frac{2m}{er}}} dr^{2} + r^{2} d \Omega^{2},     
 \label{2.4}
 \end{equation}  
 with a horizon at $r_{H} = 1/x_{H} = 2m/e$. We have always $f(r) \equiv -g_{tt} >0$ (no signature flip) and, in addition, $f(r)$ tends to unity both when $r \rightarrow 0$ and $r \rightarrow \infty$. From the shape of the curve $f(r)$ we could conclude that, close to the origin $r = 0$, $f(r)$ resembles the parabolic form of the static deS spacetime. The metric coefficient $f(r)$ has two inflexion points at $r = (m/e)(2 \pm \sqrt{2})$, located symmetrically w.r.t. $r_{H}$. 
   
 Let us take now a static observer characterized by the velocity vector field
      \begin{equation}
   u^{b} = \left(\frac{1} {\sqrt{1 - \frac{2m}{r} e^{-\frac{2m}{er}}}}, 0, 0, 0\right)    
 \label{2.5}
 \end{equation} 
 where $b$ labels $(t,r,\theta,\phi)$. From (2.5) we find the acceleration 4 - vector to be 
  \begin{equation}
  a^{b} = \left(0, \frac{m(1 - \frac{r_{H}}{r})} {r^{2}} e^{-\frac{r_{H}}{r}}, 0, 0, \right).    
 \label{2.6}
 \end{equation} 
 One sees that $a^{r}$ is vanishing when $r \rightarrow 0$ and at the horizon $r_{H} = 2m/e$. It is negative for $r < r_{H}$ and positive for $r > r_{H}$. That means the gravitational field is repulsive for $r < r_{H}$ and attractive for $r > r_{H}$. This is in accordance with our interpretation of the deS-like metric close to the origin. 
 
 From (2.6) one easily  obtains the invariant acceleration
       \begin{equation}
 a \equiv  \sqrt{ a^{b}a_{b}} = \frac{m|1 - \frac{r_{H}}{r}|e^{-\frac{r_{H}}{r}}}{r^{2} \sqrt{1 - \frac{2m}{r} e^{-\frac{r_{H}}{r}}}}.   
 \label{2.7}
 \end{equation} 
 Once we have localized the BH horizon, we look now for its surface gravity $\kappa$. Eqs. (2.4) and (2.7) yield
  \begin{equation}
 \kappa =  \sqrt{ a^{b}a_{b}} \sqrt{-g_{tt}}|_{r = r_{H}} = 0.   
 \label{2.8}
 \end{equation} 
In other words, the Hawking BH temperature is vanishing (no Hawking radiation). Therefore, our case corresponds to an extremal BH \cite{NSS} and we are left with a ''frozen'' horizon. In the Bonanno - Reuter model, $T_{H} = 0$ means the BH evaporation stops and the mass $m$ is critical ( a soliton-like remnant is formed \cite{BR}). 

In spite of the lack of a non-null surface gravity, the invariant acceleration $a$ from (2.7) is finite on the horizon. Taking the limit $r \rightarrow r_{H}$ in (2.7), we get 
 \begin{equation}
   \sqrt{ a^{b}a_{b}}|_{H} = \frac{e\sqrt{2}}{4m}
 \label{2.9}
 \end{equation} 
 It is worth noting that the expression of the invariant acceleration on the horizon resembles the surface gravity $\kappa_{S} = 1/4m$ for the KS spacetime. Concerning the curvature invariants, while the Ricci scalar $R^{a}_{~a} = (8m^{3}/e^{2}r^{5})e^{-2m/er}$ is everywhere nonsingular, the Kretschmann scalar vanishes at $r_{H}/2$ and somewhere between $r_{H}/30$ and $r_{H}/28$.
 
 \section{Anisotropic stress tensor}
 Let us see now what are the sources of the spacetime (2.4), namely what energy-momentum tensor do we need on the r.h.s. of Einstein's equations $G_{\mu \nu} = 8\pi T_{\mu \nu}$ in order (2.4) to be an exact solution. One finds, by means of the software package Maple - GRTensorII, that
   \begin{equation}
   \begin{split}
  T^{t}_{~t} = -\rho = \frac{- m^{2}}{2\pi er^{4}} e^{-\frac{2m}{er}},~~~ T^{r}_{~r} = p_{r} = - \rho,\\ T^{\theta}_{~\theta} = T^{\phi}_{~\phi} = p_{\theta} = p_{\phi} =  \frac{ m^{2}}{2 \pi e r^{4}} \left(1 - \frac{m}{er}\right) e^{-\frac{2m}{er}}.
  \end{split}
\label{3.1}
\end{equation}
We notice firstly that $\rho > p_{\theta}$ always and $p_{r} = - \rho$, as for the deS geometry. Nevertheless, the fluid is anisotropic since $p_{r} \neq p_{\theta} = p_{\phi}$. The energy density and all pressures are non-singular at $r = 0$ and when $r \rightarrow \infty$ (where, actually, they vanish). Moreover, $\rho$ is positive for any $r$ and the weak energy condition is fulfilled. However, the strong energy condition is not satisfied for $r < r_{H}/2$, where $\rho + \Sigma p_{i} = 2p_{\theta} < 0 ~(i = 1,2,3)$. The maximum value $\rho_{max} = 8/\pi m e^{2}$ is reached in the BH interior, more precisely at $r = r_{H}/4$. For $r >> r_{H}$, $\rho(r)$ tends to zero (the derivative $d\rho/dr$ depends on the BH mass, too). For example, on the surface of a star with $M_{S} \approx 10^{33} g$ and $r_{S} \approx 10^{11} cm$, one obtains $\rho \approx GM_{S}^{2}/2\pi er_{S}^{4} =  10^{14} erg/cm^{3}$ (the exponential function is practically unity). The conclusion is that, even though we are far away from the hypothetical star's horizon, $\rho$ is not negligible. In other words, the modified KS metric (2.4) keeps traces of the stress tensor even far from its horizon.

Let us observe that, for $r >> r_{H}, \rho$ may be written as $\rho \approx (1/2\pi) (m/r^{2})^{2}$ and $a \equiv \sqrt{ a^{b}a_{b}} \approx m/r^{2}$, if we also neglect $r_{H}/r$ w.r.t. unity. Therefore, we get $\rho \propto a^{2}$, a well-known result from Newtonian gravitation and electrostatics, where the energy density is proportional with the electric field squared. Indeed, for a point charge $q$ at rest, we have \cite{PI}
 \begin{equation}
 T^{a}_{(e)b} = \frac{q^{2}}{8\pi r^{4}} (-1, -1, 1, 1),
 \label{3.2}
 \end{equation} 
where $T^{a}_{(e)b}$ is the Maxwell stress tensor representing the static electric field of the point charge of strength $q$, located at the origin. It is easy to see that, far from the horizon $r = r_{H}$, the components of the stress-energy tensor (3.1) are similar with those of (3.2), provided we replace $2m/\sqrt{e}$ with $q$.

 So we may read $\rho$ as the energy density of the gravitational field sourced by the mass $m$. If the central mass is not a BH (the case of a star) we could apply Birkhoff's theorem and put all the star mass inside its gravitational radius, with no change of the gravitational field outside. We mention also that the expression for $\rho$ remains valid for the universe as a whole. Indeed, taking $r = R_{U}$ and $m = M_{U}$ and keeping in mind that $M_{U}/R_{U} \approx 1$, namely $R_{U} \approx R_{H}$, we get $\rho = 1/2\pi e^{2}R_{U}^{2} \approx 10^{-30} g/cm^{3}$, which is the value from Cosmology.

The paper by Xiang et al. \cite{XLS} (and even the previous ones) set their parameter $\alpha$ at a value of the order of the Planck length squared, $l_{P}^{2}$. Therefore, they obtain $T^{t}_{~t} = l_{P}^{2} \Phi/2\pi r^{4}$, a much smaller value compared to ours. 

As far as the transversal pressures are concerned, they vanish at $r = r_{H}/2$ and have two extrema: a minimum at $r_{1} = (7 - \sqrt{17})r_{H}/16$ and a maximum at $r_{2} = (7 + \sqrt{17})r_{H}/16$. As the other physical parameters analyzed before, our $p_{\theta}$ and $p_{\phi}$ vanish when $r \rightarrow 0$ or $r \rightarrow \infty$. In addition, they equal the energy density for $r >> r_{H}$.

 We intend to write now the Poisson equation corresponding to the classical potential $\Phi(r)$ from (2.2), with $k = 2m/e$. We have
  \begin{equation}
 \nabla^{2}\Phi(r) \equiv \frac{1}{r^{2}} \frac{d}{dr}\left(r^{2} \frac{d\Phi}{dr}\right) = \frac{4m^{2}}{er^{4}}\left(1 - \frac{m}{er}\right)e^{-\frac{2m}{er}}.   
 \label{3.3}
 \end{equation}  
It can be written as $\nabla^{2}\Phi(r) = 4\pi \cdot 2p_{\theta}$, which is different from $\nabla^{2}\Phi(r) = 4\pi \rho$, the standard Poisson equation. In our view, the discrepancy comes from the role played by the pressures in General Relativity. As Padmanabhan has shown \cite{TP1} we ought to pass from $\rho$ to $\rho_{Komar} = \rho + 3p$ (for a perfect fluid). Nevertheless, our fluid is anisotropic so that we must replace $\rho + 3p$ with $\rho + p_{r} + p_{\theta} +p_{\phi} = 2p_{\theta}$, as it verifies for (2.5) ($2p_{\theta}$ appears as the field source).

\section{Tolman - Komar energy}
Our next task is to evaluate the Tolman-Komar (TK) gravitational energy for the metric (2.4). It is given by \cite{TP, HC}
   \begin{equation}
W = 2 \int(T_{ab} - \frac{1}{2} g_{ab}T^{c}_{~c})u^{a} u^{b} N\sqrt{\gamma} d^{3}x ,
\label{4.1}
\end{equation}
and is measured by a static observer with the velocity field $u^{a}$. $N$ in (3.1) is the lapse function and $\gamma$ is the determinant of the spatial 3-metric.With $u^{a}$ from (2.6), Eq. (4.1) yields
 \begin{equation}
 W = \int^{r}_{0}\frac{4m^{2}}{er^{4}}\left(1 - \frac{2m}{er}\right)e^{-\frac{2m}{er}}r^{2}dr.  
\label{4.2}
\end{equation}
With the substitution $x = 1/r$, one finally finds that
 \begin{equation}
 W = m\left(1 - \frac{r_{H}}{r}\right) e^{-\frac{r_{H}}{r}}
\label{4.3}
\end{equation}
It is clear from (4.3) that $W$ tends to zero when $r \rightarrow 0$. However, $W = m$ at infinity, as expected. In other words, the total TK energy of the anisotropic fluid equals the rest mass of the central body. A similar result has been obtained by Bonanno and Reuter \cite{BR} where the total energy of their fluid is identified with the extremal BH mass $M$. Nevertheless, our gravitational energy (4.3) becomes negative for $r < r_{H}$ and presents a minimum $W_{min} = -m/e^{2}$ at $r = r_{H}/2$, which seems to be a steady state. This is rooted from the negative pressures contribution. 

One may be interested to compute the ADM mass for the asymptotically flat spacetime (2.4). We have
 \begin{equation}
 W = \int^{\infty}_{0} 4\pi r^{2}\rho(r) dr = m
\label{4.4}
\end{equation}
This is another confirmation of the correctness of our interpretation of the stress tensor (3.1). Even though $W$ from (4.3) and the BH temperature vanish on the horizon, the horizon entropy $S = |W|/2T$ \cite{TP} is finite. Indeed, one obtains 
\begin{equation}
S_{H} = \left(\frac{|W|}{2T}\right)_{H} = \left(\frac{|W|\pi r^{2}}{\kappa}\right)_{H} = \pi r_{H}^{2} = \frac{4\pi m^{2}}{e^{2}} 
\label{4.5}
\end{equation}
 and we see that the relation $S_{H} = A_{H}/4$ is observed.

\section{Horizon stress tensor}
Let us find now the expression of the surface energy-momentum tensor on the horizon $r_{H} = 2m/e$. We must have a nonzero energy on this hypersurface due to the jump of its extrinsic curvature $K_{ab}$ when the horizon is crossed. To see this, we write down the expression of the extrinsic curvature tensor \cite{KP, HC, HC1} 
 \begin{equation}
 K_{ab} = -\frac{f'}{2\sqrt{f}} u_{a}u_{b} + \frac{\sqrt{f}}{r} q_{ab},
\label{5.1}
\end{equation}
where $f'\equiv df(r)/dr$, $u_{a} = (\sqrt{f}, 0, 0, 0)$ (obtained from (2.5)) is the normal to $t = const.$ hypersurface, $h_{ab} = g_{ab} - n_{a}n_{b}$ represents the induced metric on a $r = const.$ surface with $n_{a} = (0, 1/\sqrt{f},0, 0)$ its normal vector and $q_{ab} =  h_{ab} + u_{a}u_{b}$ is the induced metric on the 2-surface of constant $t$ and $r$ \cite{KP}. Eq. (5.1) yields
 \begin{equation}
K_{tt} = -\frac{f'}{2} \sqrt{f},~~~K_{\theta \theta} = \frac{K_{\phi \phi}}{sin^2 \theta} = r \sqrt{f},~~~ K \equiv h^{ab}K_{ab} = \frac{f'}{2\sqrt{f}} + \frac{2\sqrt{f}}{r}
\label{5.2}
\end{equation}
We suppose the horizon is a membrane with a surface stress tensor $S_{ab}$ of a perfect fluid form
 \begin{equation}
 S_{ab} = \rho_{H} u_{a}u_{b} + p_{H} q_{ab},
\label{5.3}
\end{equation}
with $\rho_{H}$ - the surface energy on $r = r_{H}$ and $p_{H}$ - the surface pressure. $S_{ab}$ is obtained from the Lanczos equation
 \begin{equation}
 8\pi S_{ab} = [K h_{ab} - K_{ab}],
\label{5.4}
\end{equation}
where $[K_{ab}] = K_{ab}^{+} - K_{ab}^{-}$ is the jump of the extrinsic curvature when the horizon is crossed, from $r > r_{H}$ to $r < r_{H}$.

The first Israel junction condition is automatically satisfied as we use same coordinates on either side of the horizon $r_{H}$. In addition, the metric on the horizon is simply \cite{HC1} $ds^{2}_{H} = r_{H}^{2} d\Omega^{2}$, when we put $r = r_{H}$ in (2.4). We shall determine $\rho_{H}$ and $p_{H}$ from (5.3) so that the second junction conditions (5.4) to be obeyed.

We write now explicitly the expression of the mean curvature 
 \begin{equation}
 K = \frac{m(1 - \frac{r_{H}}{r})e^{-\frac{r_{H}}{r}}}{r^{2} \sqrt{1 - \frac{2m}{r} e^{-\frac{r_{H}}{r}}}} + \frac{2}{r} \sqrt{1 - \frac{2m}{r} e^{-\frac{r_{H}}{r}}}  
 \label{5.5}
 \end{equation} 
taken at $r = r_{H}$. One sees that the second term from the r.h.s. of (5.5) has no jump when the horizon is crossed. On the contrary, for the first term both the numerator and denominator are vanishing at $r = r_{H}$ and, therefore, we must consider the limit at $r \rightarrow r_{H}$. But the side limits are different (they have opposite values). Evaluating the limits, one obtains $K_{+} = -K_{-} = e\sqrt{2}/4m$. This is exactly the value of the invariant acceleration $a$ from (2.9) or the values of normal components $(a^{b}n_{b})_{+} = - (a^{b}n_{b})_{-}$ on the horizon. From (5.4) we have $S_{tt} = 0$ and $S_{\theta \theta} = m/2\pi e\sqrt{2}$. Therefore, from (5.3) one obtains
 \begin{equation}
 \rho_{H} = 0,~~~p_{H} = \frac{e}{8\pi \sqrt{2}m}.
 \label{5.6}
 \end{equation} 
A vanishing value of the surface energy $\rho_{H}$ has been also obtained by Kolekar et al. \cite{KP} (see also \cite{HC, HC1}). However, their surface pressure $p_{H}$ was divergent because only the denominator of the ratio $f'/2\sqrt{f}$ is null on the horizon. Nevertheless, $(f'/2\sqrt{f})_{H}$ is finite in the present work and, as a consequence, $p_{H} = -\sigma_{H}$ from (5.6) is finite. In terms of the proper acceleration, (5.6) gives us $p_{H} = a_{H}/4\pi$.

Let us check now whether the relativistic Young-Laplace equation \cite{CDR}
 \begin{equation}
 p_{r} = \sigma K
 \label{5.7}
 \end{equation} 
is observed at $r = r_{H}$. Here $p_{r}$ is the bulk pressure from (3.1) (note that $p_{H}$ from (5.6) has units $erg/cm^{2}$). Keeping in mind that $p_{r}$ (at $r = r_{H}$) = - ($m^{2}/2\pi er^{4})e^{-\frac{r_{H}}{r}} |_{H} = -e^{2}/32\pi m^{2},\sigma_{H} =-p_{H} =-e/8\pi \sqrt{2}m$ and $K = e\sqrt{2}/4m$, we notice that (5.7) is obeyed.

We remind here that the horizon is viewed as a spherical membrane with a negative surface tension $\sigma_{H}$, to whom the Young-Laplace equation may be applied. Let us estimate the order of magnitude of $p_{H}$ for a Solar mass black hole. Taking $m = 2.10^{33}$g and introducing the universal constants $G$ and $c$, we get $p_{H} = (c^{6}/G^{2})(e/8\pi \sqrt{2}m) \approx 10^{42} erg/cm^{2}$. 

\section{Near-horizon approximation}
We look now for the near horizon version of the spacetime (2.4), following an analogy with the standard Schwarzschild metric, to which the near horizon metric is approximately Rindler. We develop the function $f(r)$ in a power series around $r = r_{H}$, up to the 2nd order
 \begin{equation}
f(r) \approx f(r_{H}) + (r - r_{H}) f'(r_{H}) + \frac{1}{2} (r - r_{H})^{2} f''(r_{H})
 \label{6.1}
 \end{equation} 
Keeping in mind that
 \begin{equation}
f''(r) = - \frac{4m}{r^{3}}\left(1 - \frac{4m}{er} + \frac{2m^{2}}{e^{2}r^{2}}\right)e^{-\frac{2m}{er}},
 \label{6.2}
 \end{equation} 
$f(r_{H}) = 0$ and $f'(r_{H}) = 0$, (6.1) yields
 \begin{equation}
f(r) = \frac{1}{2} \frac{(r - r_{H})^{2}}{r_{H}^{2}},
 \label{6.3}
 \end{equation} 
where the relation $f''(r_{H}) = 1/r_{H}^{2}$ has been used. 

In terms of the invariant acceleration on the horizon, we have $r_{H}^{2} = 1/2a_{H}^{2}$ and, therefore, the near horizon metric appears as
   \begin{equation}
  ds^{2} = - a_{H}^{2} (r - r_{H})^{2} dt^{2} + \frac{ dr^{2}}{a_{H}^{2} (r - r_{H})^{2}} + r_{H}^{2} d \Omega^{2}. 
 \label{6.4}
 \end{equation}
We notice that the two-dimensional $r - t$ sector of (6.4) resembles the Dadhich ''free of centrifugal acceleration'' geometry \cite{ND} (see also \cite{HC2} where the Dadhich constant $k$ has been interpreted as an acceleration squared), if we perform the translation $r - r_{H} \rightarrow r'$. The full geometry (6.4) is equivalent to the near-horizon approximation of the critical quantum BH \cite{BR} (Bonanno and Reuter metric (4.23), where $r_{cr}$ plays the role of our $r_{H}$). It resembles the Robinson-Bertotti (RB) space for the product of a two-dimensional $AdS_{2}$ space with a two-sphere, namely $AdS_{2}x S^{2}$. However, (6.4) is not exactly the RB metric because the $S^{2}$ curvature is $r_{H}$ but the $AdS_{2}$ curvature is given by $r_{H}\sqrt{2}$. Consequently, (6.4) is not conformally flat (see also \cite{SF}).

It would be interesting to see what stress tensor $\bar{T_{ab}}$ should be placed at the r.h.s. of Einstein's equations for to render the metric (6.4) an exact solution. One finds that
    \begin{equation}
		\bar{\rho} = - \bar{p}_{r} = 2\bar{p}_{\bot} = \frac{1}{8\pi r_{H}^{2}}
 \label{6.5}
 \end{equation}
Let us observe that the above values coincide with those from Eqs. (3.1) , taken at the horizon $r_{H} = 2m/e$. The radial acceleration is now $\bar{a}^{r} = (r - r_{H})/2r_{H}^{2}$, with $\sqrt{\bar{a}^{b}\bar{a}_{b}} = e\sqrt{2}/4m$, which is exactly its value on the horizon.

We compute now, for completeness, the jump of the extrinsic curvature when the horizon $H$ is crossed, with $f(r)$ from (6.3). One must remind that we have now $g_{\theta \theta} = r_{H}^{2} = const.$ and, therefore, the components of the extrinsic curvature of hypersurfaces of constant $r$ are no longer given by (5.1) and (5.2). Indeed, the connection coefficient $\Gamma_{\theta \theta}^{r} = 0$ and, therefore, $\bar{K}_{\theta \theta} = 0$.  In addition, one obtains
    \begin{equation}
		\bar{K}_{tt} = - \Gamma_{tt}^{r} n_{r} = - \frac{\sqrt{2}(r - r_{H})^{3}}{4r_{H}^{3} |r - r_{H}|}.
 \label{6.6}
 \end{equation}
Eq. (6.6) gives us now
    \begin{equation}
		\bar{K} = \bar{K}^{a}_{a} = \bar{K}^{t}_{t} =   \frac{\sqrt{2}(r - r_{H})}{2r_{H} |r - r_{H}|}.
 \label{6.7}
 \end{equation}
The jump of $\bar{K}$ will be $[\bar{K}] = \frac{\sqrt{2}}{r_{H}}$, exactly the previous value obtained from (5.2). As a consequence, we get again $\rho_{H} = 0$ and $p_{H} = e/8\pi \sqrt{2}m$. It is worth noting that, for the geometry (6.4), $K_{ab}$ from (5.1) contains only the 1st term and, therefore, $\bar{K} = \bar{K}^{t}_{t}$.

\section{Conclusions}
The late stages of the evaporation process of a BH and the role of quantum gravitational effects sparked much interest in the last decade. The problem is whether the evaporation stops and a remnant arises.

We looked in this paper for a solution to the problem by means of classical arguments, with no use of quantum effects. As many of the previous authors, we proposed a modification of the KS geometry to render the physical parameters finite at $r = 0$ and to get rid of any singularities. For the spacetime (2.4) to be an exact solution of Einstein's equations, the field must be sourced by an anisotropic fluid with negative radial pressure $p_{r} = -\rho$. While $\rho$ and $p_{r}$ preserve their sign for any $r$, the transversal pressures fluctuate for $r < r_{H}$ but for $r >> r_{H}$ equal the energy density. In addition, $\rho \propto a^{2}$ far from the horizon, in accordance with similar dependence from Newtonian gravitation and electrostatics. We found that the gravitational field (2.4) is repulsive for $r < r_{H}$ and attractive for $r > r_{H}$. Even though our metric exhibits a horizon at $r_{H} = 2m/e$, the surface gravity $\kappa$ vanishes and, therefore, $T_{BH} = 0$ (the horizon is frozen and the BH becomes extremal). The TK energy $W$ was calculated by means of Padmanabhan's formula and proves to change its sign at $r = r_{H}$ and to equal the ADM mass at infinity. Even though the BH temperature and TK energy are vanishing on the horizon, the horizon entropy $S_{H}$ is finite and equals $A_{H}/4$, as it should be. Our modification of the KS metric suggest that, outside a central body with spherical symmetry we have gravitational energy density given by $T^{a}_{~b}$ from (3.1), with the global energy $W = m$. Moreover, due to a jump of the extrinsic curvature when the horizon is crossed, a surface pressure $p_{H} = e/8\pi \sqrt{2}m$ is present but the surface energy density is vanishing.


\begin{thebibliography} {18}

\bibitem{ID}
I. Dymnikova, Gen. Relat. Grav. 24, 235 (1992).
\bibitem{DK}
I. G. Dymnikova, Phys. Lett. B472, 33 (2000) (arXiv: gr-qc/9912116); I. Dymnikova and M. Korpusik, Entropy 13, 1967 (2011). 
\bibitem {BR}
A. Bonanno and M. Reuter, Phys. Rev. D62, 043008 (2000) (arXiv: hep-th/0002196).
\bibitem{NSS}
P. Nicolini, A. Smailagici and E. Spallucci, Phys. Lett. B632, 547 (2006) (arXiv: gr-qc/0510112).
\bibitem{PN}
P. Nicolini, J. Phys. A38, L631 (2005) (arXiv: hep-th/0507266); Int. J. Mod. Phys. A24, 1229 (2009) (arXiv: 0807.1939 [hep-th]). 
\bibitem{MKP}
Y. S. Myung, Y.-W. Kim and Y.-J. Park, Phys. Lett. B656, 221 (2007) (arXiv: gr-qc/0702145).
\bibitem {SAH}
S. A. Hayward, Phys. Rev. Lett. 96, 031103 (2006) (arXiv: gr-qc/0506126).
\bibitem {XLS}
L. Xiang, Y. Ling and Y. G. Shen, arXiv: 1305.3851 [gr-qc].
\bibitem{TP1}
T. Padmanabhan, Phys. Rev. D81, 124040 (2010) (arXiv: 1003.5665 [gr-qc]); Mod. Phys. Lett. A25, 1129 (2010) (arXiv: 0912.3165 [gr-qc]).
\bibitem{PI}
E. Poisson, W. Israel, Phys. Rev. D41, 1796 (1990).
\bibitem{TP}
T. Padmanabhan, Class. Quantum Grav. 21, 4485 (2004) (arXiv: gr-qc/0308070).
\bibitem{HC}
H. Culetu, Int. J. Mod. Phys.: Conf. Series, 3, 455 (2011) (arXiv: 1101.2980 [gr-qc]); arXiv: 1304.5386 [gr-qc].
\bibitem{KP}
S. Kolekar and T. Padmanabhan, Phys. Rev. D85, 024004 (2011) (arXiv: 1109.5353 [gr-qc]); S. Kolekar, D. Kothawala and T. Padmanabhan, Phys. Rev. D85, 064031 (2012) (arXiv: 1111.0973 [gr-qc]).
\bibitem{HC1}
H. Culetu, Phys. Lett. A376, 2817 (2012).
\bibitem{CDR}
V. Cardoso, O. Dias and J. Rocha, JHEP 1001, 021 (2010) (arXiv: 0910.0020 [hep-th]).
\bibitem{ND}
N. Dadhich, arXiv: 1209.2144 [gr-qc].
\bibitem{HC2}
H. Culetu, arXiv: 1303.7376 [gr-qc].
\bibitem{SF}
S. Ferrara, arXiv: hep-th/9701163.




\end{thebibliography}
\end{document}